# Efficacy of static analysis tools for software defect detection on open-source projects

Jones Yeboah

School Of Information Technology

University Of Cincinnati, USA

Email:

yeboahjs@mail.uc.edu

Saheed Popoola

School Of Information Technology

University Of Cincinnati, USA

Email:

saheed.popoola@uc.edu

0000-0002-9602-6322

*Abstract*— In software practice, static analysis tools remain an integral part of detecting defects in software and there have been various tools designed to run the analysis in different programming languages like Java, C++, and Python. This paper presents an empirical comparison of popular static analysis tools for identifying software defects using several datasets using Java, C++, and Python code. The study used popular analysis tools such as SonarQube, PMD, Checkstyle, and FindBugs to perform the comparison based on using the datasets. The study also used various evaluation metrics such as Precision, Recall, and F1-score to determine the performance of each analysis tool. The study results show that SonarQube performs considerably well than all other tools in terms of its defect detection across the various three programming languages. These findings remain consistent with other existing studies that also agree on SonarQube being an effective tool for defect detection in software. The study contributes to much insight on static analysis tools with different programming languages and additional information to understand the strengths and weaknesses of each analysis tool. The study also discusses the implications for software development researchers and practitioners, and future directions in this area. Our research approach aim is to provide a recommendation guideline to enable software developers, practitioners, and researchers to make the right choice on static analysis tools to detect errors in their software codes. Also, for researchers to embark on investigating and improving software analysis tools to enhance the quality and reliability of the software systems and its software development processes practice.

## I. Introduction

Software defects have several negative consequences and when not managed properly can affect project delivery deadlines and development costs. It can even impact customers with poor user experience and security flaws in the software system [1]. For software development teams to mitigate these risks, there is the need to use static analysis tools early in the development process to detect defects in codes [2].

As the programming world has evolved, static analysis has become a valuable technique in software development for identifying errors without running the software system [3]. This technique is gaining in popularity because it enables error detection and correction early in the development process, resulting in low-cost and efficient error detection and correction [2]. Various static analysis tools are available to analyze various aspects of software systems. These tools are very effective at detecting errors, but their effectiveness depends on the type of error and the programming language used [4].

Static analysis tools aid in analyzing software codes without running it with the aim of identifying common programming errors, security flaws or other issues that can cause problems with the performance of the software system [3]. These tools can assist software developers to identify common errors more quickly before it becomes a serious issue thereby saving project delivery timelines and reducing development costs.

Also, it is imperative to note that while many static analysis tools are available on the software market, each tool has its own strengths and weaknesses. Some specific tools work effectively on specific programming languages or projects while others are easy to use in a more user-friendly environment without much technical expertise and have high accuracy rates [4].

Specifically, we will investigate the following research question:

1. How accurately do different static analysis tools identify common software defects?
2. What is the significant difference in the effectiveness of each tool capability to detect software defects?

For this research question to be answered, we will compare four static analysis tools using a common dataset. Our study will provide a better understanding of the strengths and weaknesses of the various static analysis tools and how that influences the decision-making of software developers to make the right choice in choosing a specific tool for their software code analysis in their project.

The paper comprises five sections: In section I we introduced static analysis tools. In section II, we provided information about the related works done in static analysis tools, and their software defects capabilities. In section III, we describe the approach to the research methodology for our study which comprises the evaluation metrics used as well as the datasets and tools used. In section IV we present the analysis of our results. In section V, we discussed the summary of the findings and future directions for the study. Finally, in section VI, we conclude with the main contribution made by the study.

## II. RELATED WORK

### A. Contribution of this Study

However, studies on the effectiveness of other types of static analysis tools in identifying software errors are limited, as most of these studies focus primarily on comparing troubleshooting tools [12]. Furthermore, these studies mainly used proprietary software systems, so there is uncertainty as to whether the results are applicable to open source software systems.

This study aims to compare four static analysis tools with a focus on much larger scale and diverse datasets that comprises of 50 software datasets. We hope this study will provide a meaningful insight into each tool based on its strengths and weaknesses. We will also test and evaluate the tools with different projects written in multiple programming languages to provide us with a generalization of the results. Also, to evaluate how easy it is to use the tools to perform the analysis to understand how software developers can use them for practice. Our research study will provide the researchers with a comprehensive finding on static analysis tools and assist software developers and practitioners in making the correct decisions when selecting a static analysis tool for software development projects and its practices.

## III. METHODOLOGY

### B. Static Analysis Tools

The four static analysis tools to compare in this study:

SonarQube [1] is a popular open-source tool for analyzing static code quality. Sonarcloud.io offers it as a service or it can be downloaded and installed on a private server. A number of metrics are computed by SonarQube, such as lines of code and the complexity of the code, as well as the compliance of the code with a specific set of coding rules that have been developed for most popular programming languages. The tool reports an 'issue' if the analyzed source code violates a coding rule. Remediation effort refers to the amount of time required to resolve these issues.
SonarQube includes rules for reliability, maintainability, and security. A reliability rule, also known as a bug, creates quality issues that indicate something is wrong with the code, which will soon be reflected in a bug. Code smells are defined as maintenance-related issues in the code that decrease the readability and modifiability of the code. In the category code smells, SonarQube includes some of the code smells [13].

FindBugs[2] is a static analysis tool for evaluating Java code, more specifically Java bytecode. Even though the tool analyzes bytecode, it can pinpoint the exact location of an issue if the source code is also provided. The tool analysis is performed using a GUI [3], which is accessed through a command line interface. A bug pattern is detected by the tool as part of the analysis. According to FindBugs, the bug patterns arise from the following factors: difficult language features, misunderstood API features, misunderstood invariants when code is modified during maintenance and simple mistakes. [4][5] As a result, these bug patterns are categorized into nine different categories: bad practice, correctness, experimental, internationalization, malicious code vulnerability, multithreaded correctness, performance, security, and dodgy code. Furthermore, the bug patterns are ranked from 1 to 20. In SourceForge report, rank 1–4 is the scariest group, rank 5–9 is the scary group, rank 10–14 is the troubling group, and rank 15–20 is the concern group. [6]

Checkstyle[7] is an open-source tool for evaluating Java code quality. A command line tool or a side feature of Ant is used to perform the analysis. According to a set of checks, Checkstyle evaluates code conforming to a certain coding standard. In Checkstyle, there are two sets of style configurations for standard checks: Google Java Style[8] and Sun Java Style.[9] It is also possible to create customized configuration files according to the user's preferences [10] in addition to the standard checks. Annotations, block checks, class design, coding, headers, imports, Javadoc comments, metrics, miscellaneous, modifiers, naming conventions, regexp, size violations, and whitespace are among the 14 categories. In addition, violations of the checks are classified into two severity levels: error and rule, [11] with the first indicating actual problems and the second indicating potential problems to be resolved.

---

[1] http://www.sonarsource.org/
[2] http://findbugs.sourceforge.net
[3] http://findbugs.sourceforge.net/manual/gui.html
[4] http://findbugs.sourceforge.net/findbugs2.html
[5] http://findbugs.sourceforge.net/factSheet.html
[6] http://findbugs.sourceforge.net/bugDescriptions.html
[7] https://checkstyle.org
[8] https://checkstyle.sourceforge.io/google_style.html
[9] https://checkstyle.sourceforge.io/sun_style.html
[10] https://checkstyle.sourceforge.io/index.html

PMD[12] is primarily used to evaluate Java and Apex, but it can also be applied to six other programming languages. Through the command line, the tool's binary distributions are used to conduct the analysis. PMD assesses code quality according to a set of rules, such as unused variables, empty catch blocks, unnecessary object creation, and more. A total of 33 different rule set configurations[13] are available for Java projects. It is also possible to customize the rule sets according to the preferences of the user.[14] In total, there are eight categories of rules: best practices, code style, design, documentation, error prone, multi-threading, performance, and security. Further, violations of the rules are ranked on a priority scale from 1 to 5, with 1 being the most serious and 5 being the least serious.[15]
A description of the priority guidelines for default and custom-made rules can be found in the PMD project documentation.

In general, SonarQube, FindBugs, CheckStyle and PMD. These tools were chosen due to their popularity in both academia and the software industry and have been used in detecting various software defects in different programming languages.

The FindBugs tools are widely used to analyze Java-based software projects to assist in detecting bugs, ensuring correctness of codes, security and improving performance in the software system. PMD as a tool also works with Java dependent software projects with the aim to detect overall complex codes, check coding writing style and standard best practices. SonarQube is a tool that performs software projects dependent on different programming languages such as C/C++ and Python.

*C. Dataset*

To test and evaluate the performance of each tool in our study. We selected open-source projects from different dimensions such as scientific computing, web development and system programming from GitHub. The list can be found here. Also, the projects selected were written in Java, C/C++ and Python to allow evaluation of the various tools based on their capabilities on different programming languages. We chose a total of fifty (50) projects which come with various sizes and complexity to ensure we have diverse codebases to evaluate the tools.

*D. Metrics*

The following metrics were used to evaluate the performance of our static analysis tools:

F-measure, precision, and recall. Basically, precision is how many true positives are in comparison with how many predicted positives there are. You can calculate recall by dividing true positives by the sum of true positives F-measure calculates the harmonic mean of precision and recall.

F-measure = 2 × (precision × recall) / (precision + recall)

Machine learning and information retrieval use these metrics to evaluate binary classification systems. In our case, True Positive (TP) happens when the tool correctly identifies the error, False Positive (FP) happens when it doesn't find the error, False Negative (FN) happens when true error exists and true negative (TN) occurs when the tool correctly determines that there is no error.

*E. Experimental Design*

The experiment will be in four (4) groups with each 50 projects assigned randomly to a group and each group using a different tool. This approach is to allow all projects to be tested equally with all the four static analysis tools. In this case, any difference in the performance of each tool is attributed to the tool itself but not to the differences in the software project used during the testing. A cross-validation approach will be adopted and that presents the results in a more reliable and robust manner and go ahead to perform other statistical analysis to determine if there are any significant differences in the performance of the tools.

## IV. RESULTS

*F. Performance Comparison*

Our study evaluated the performance of the four static analysis tools that comprise FindBugs, PMD, CheckStyle, and SonarQube on datasets from fifty (50) open-source projects written in multiple programming languages namely Java, C/C++ as well as Python. Table 1 below represents the results of the study that depicts the performance of each tool based on their precision, recall and F1-score.

Table 1: Performance comparison of four (4) static analysis tools (**RQ₁**)

| Tool | Precision | Recall | F1-score |
|---|---|---|---|
| FindBugs | 0.78 | 0.82 | 0.80 |
| PMD | 0.71 | 0.76 | 0.73 |
| Checkstyle | 0.69 | 0.71 | 0.70 |
| SonarQube | 0.83 | 0.87 | 0.85 |

---

[11]https://checkstyle.sourceforge.io/checks.htm
[12]https://pmd.github.io/latest/
[13]https://github.com/pmd/pmd/tree/master/pmd-java/src/main/resources/rulesets/java
[14]https://pmd.github.io/latest/index.htm
[15]https://pmd.github.io/latest/pmd_rules_java.html

In the Table 1 above, it clearly shows that SonarQube had the highest F1-score percentage of 85% which gives a clear indication of its effectiveness in detecting defects in the datasets used for the study, then FindBugs, PMD and CheckStyle follow in that order in terms of the level of effectiveness.

*G. Statistical Analysis(RQ$_2$)*

Statistical analysis of the results was conducted using a one-way ANOVA test approach was used to determine whether there is any significant difference in the performance of the various static analysis tools used in the study. The test results show a significant difference in the F1-score among the static analysis tools ($F (3, 196) = 4.63$, $p < 0.05$). These results are derived from a statistical analysis test. The F-value of 4.63 provides the variation ratio between the tools used for the study and the variation within the four groups. (3 and 196) refer to the degree of freedom in this regard as shown in parenthesis. In this test, the p-value is simply lower than 0.5 which signifies that FI-score shows the statistical significance difference between the static analysis tools. We can also observe that the difference between the tools is not likely by chance only.

This analysis shows the significant difference in the effectiveness of each tool capability to detect software defects. Other further test conducted with the aim to ascertain the specific tools that are significantly different from one another using a post-hoc Turkey test approach. The test results revealed that there is a significant difference between the SonarQube as well as the CheckStyle with p-value lower than 0.05 ($p<0.05$) and between SonarQube and PMD ($p<0.05$), but the results could not find any significant differences among the SonarQube and FindBugs, or even between the other three tools. This analysis revealed that using post-hoc turkey tests the results showed that two tools performed differently than the SonarQube in a statistically different way. In other words, these tools' performance was quiet similar to SonarQube and did not show a statistically significant difference in performance in the detection of software defects.

*H. Illustration of Results*

Figure 1 below depicts the performance of the static analysis tools used in the study based on the following metrics: Precision, Recall, and F1-score. In the results, SonarQube performed as the effective tool with highest precision and recall which resulted also in high F1-score, while the other tools such as CheckStyle recorded the lowest precision, recall and F1 score.

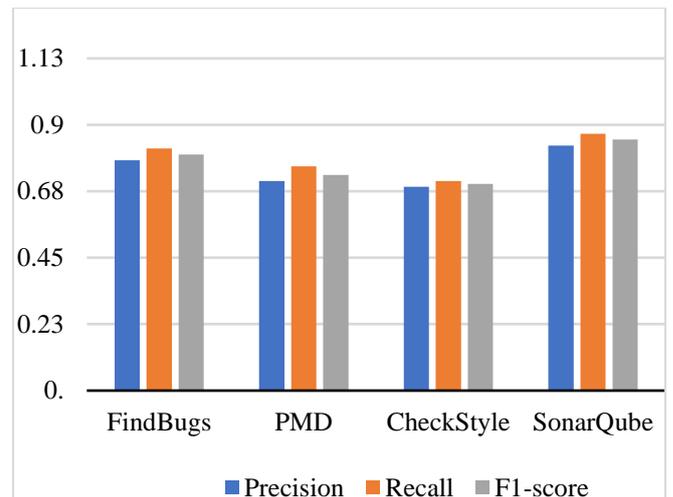

Figure 1: Performance Comparison of the Four Static Analysis Tools

## V. DISCUSSION

*I. Interpretation of Results*

The results of this study show consistency with other existing studies that also found SonarQube as a reliable static analysis tool for detecting software defects in codes [14,15]. In our research findings, SonarQube was still regarded as the most effective tool in the detection of software defects. It is followed by FindBugs, PMD, and CheckStyle in that order.

These findings in our study really support the hypothesis that SonarQube would be much more effective than the other tools used in the study. Our findings also reveal that precision and recall are essential metrics to be considered to determine the performance evaluation of static analysis tools [15].

*J. Comparison to Previous Studies*

There have been numerous studies conducted on static analysis tools which have revealed different results, which is mostly due to the programming language used or the evaluation metrics adopted to evaluate the performance of the tools [16,17]. An example of this can be a study conducted by [17] where the authors discovered that PMD and FindBugs outperformed SonarQube in the detection of certkddkain types of defects found in software projects written in the Java programming language.

*K. Strengths and Limitations*

The defining part of this study is evaluating the tools with multiple metrics to determine their performance. Also, introducing diverse datasets with different programming languages to enhance study results generalization [15]. However, the analysis focused on 50 software projects, which might not be a true reflection of all software projects. Furthermore, we only focused on open-source projects which may not provide us with generalized results for proprietary software systems [16].

*L. Practical Implications*

Our study findings provide the software developers with information that can aid them in their decision making and encouraged them to go for SonarQube or FindBugs as the main tools to detect software defects in code [15]. In terms of selecting a tool for your testing or code analysis, however, one must be guided by the needs of the project, such as its complexity, the size of the codebase, the programming languages used for the system development, and the desired balance between precision and recall.

*M. Recommendation Guidelines*

Our study results prove that adopting the right static analysis tool can improve the quality of software code and achieve a highly robust and scalable software system. Based on the study's results, we've got these guidelines.

It's important to understand that static analysis tools aren't flawless. It's still possible to improve their effectiveness. To make sure code is quality, developers shouldn't just rely on these tools.

The results of our study showed how certain tools performed better than others at identifying errors. FindBugs, for example, is better at finding errors than Checkstyle and PMD.

In our study, we identified that each of the tools utilized has their own capabilities in terms of strengths and weaknesses so it's imperative for projects to be selected based on the specific needs and requirements. This study conducted comes with its own limitations such as evaluating a specific Java project. In this regard, the results of the study cannot be applied to all software systems developed. It is important to evaluate your own project based on several conditions like the language used, the complexity of the project as well as the level of expertise of the team working on the project.

To conclude, the guidelines are very paramount and research practitioners, and the software industry can implement the framework that encourages the combination of following software quality assurance practices and the usage of satic analysis tools to achieve the maximum benefits that comes with software code analysis for detection for errors.

Furthermore, this approach will assist in the development of a more reliable and quality standard of software. It should also be considered as part of the overall code analysis, while we also focus on continuous improvement that aim to improve analysis tools to detect errors early and fix them. As a result, software systems will be more scalable, reliable, and robust.

## VI. CONCLUSION

*N. Main Findings and Implications*

The study provided us with comprehensive findings on the static analysis tools and its tool's performance in the detection of defects in software. SonarQube was regarded as the most effective tool in the detection of defects in software.

It has become easier for researchers, developers as well as testers to choose the right tool to detect software defects effectively. This will reduce development costs, prevent security flaws, and help developers meet project delivery timelines for release dates.

*O. Contributions to the Field*

This study has contributed immensely to the software engineering field with the exploration of the various static analysis tools and coming up with a result that indicate the effective static analysis tools for detecting software defects and going further to perform a comparative analysis of each tool's performance on different codebases. This study builds on existing studies by introducing the use of large and more diverse sets of codebases to perform the analysis as well as incorporating different evaluation metrics to assess each tool's performance.

*P. Limitations and Areas for Future Research*

One key limitation of this study is the use of specific tools and codebases for our analysis, which may not be a true reflection of all possible scenarios in software systems. The study also did not explore the impact of different configurations for each tool. Moving further, researchers can explore this factor as well as evaluate the static analysis tools on specific types of defects found in software. Researchers can also explore other static analysis tools such as ESLint or Infer. To further examine the generalization of the results, other studies can explore in different contexts with a focus on proprietary software systems. Finally, machine learning is an important domain that will be great to investigate to check on the various machine learning techniques that can be adopted to improve the performance of existing static analysis tools in the detection of software defects in code [18].

*Q. Final Statement*

To conclude, this study provided much insight into the potential of the use of static analysis tools in software development practices in the detection of software defects and an overview of their performance. The study results can assist developers to make decisions on the right tool to improve the quality and security of their software systems.